\documentclass[prl, a4paper, nofootinbib, twocolumn, showpacs]{revtex4}

\usepackage{amssymb, amsfonts, amsmath,dsfont,epsf,pstricks,graphicx,epsfig,epic,eepic,psfrag,latexsym, bm, verbatim, rotating,multirow}

\renewcommand{\vec}[1]{{\mathbf #1}}

\newcommand{\KK}{\mathbf{K}}
\newcommand{\dd}{\mathbf{d}}

\newcommand{\rr}{\mathbf{r}}
\newcommand{\hh}{\mathbf{h}}

\newcommand{\dsum}{\displaystyle\sum}

\psfrag{``bz''}{$b_{z}$}
\psfrag{``bx''}{$b_{x}$}
\psfrag{``by''}{$b_{y}$}
\psfrag{``hz''}{\textcolor{red}{$h_{z}$}}
\psfrag{``hx''}{\textcolor{red}{$h_{x}$}}
\psfrag{``hy''}{\textcolor{red}{$h_{y}$}}
\psfrag{``r1''}{$r_{3}$}
\psfrag{``r2''}{$r_{2}$}
\psfrag{``r3''}{$r_{1}$}
\psfrag{``h1z''}{$h_{z}\sigma^{z}_{3}$}
\psfrag{``h2y''}{$h_{y}\sigma^{y}_{2}$}
\psfrag{``h3x''}{$h_{x}\sigma^{x}_{1}$}
\psfrag{``x''}{\textcolor{blue}{x}}
\psfrag{``y''}{\textcolor{magenta}{y}}
\psfrag{``z''}{\textcolor{red}{z}}
\psfrag{``pos1''}{1}
\psfrag{``pos2''}{2}
\psfrag{``pos3''}{3}
\psfrag{``pos4''}{4}
\psfrag{``pos5''}{5}
\psfrag{``pos6''}{6}
\psfrag{``plaq''}{P}

\begin{document}
\title{Disorder in a quantum spin liquid: flux binding and local moment formation}
\author{A.\ J.\ Willans$^{1}$}
\author{J.\ T.\ Chalker$^{1}$}
\author{R.\ Moessner$^{2}$}
\date{\today}
\affiliation{$^{1}$Theoretical Physics, Oxford University, 1, Keble Road,  Oxford OX1 3NP, United Kingdom\\$^{2}$Max-Planck-Institut f\"{u}r Physik komplexer Systeme, N\"otnitzer Stra{\ss}e 38, 01187 Dresden, Germany}

\begin{abstract}
We study the consequences of disorder in the Kitaev honeycomb model, considering both site dilution and exchange randomness. We show that a single vacancy binds a flux and induces a local moment. This moment is polarised by an applied field $h$: in the gapless phase, for small $h$ the local susceptibility diverges as $\chi(h)\sim\ln(1/h)$; for a pair of nearby vacancies on the same sublattice, this even increases to $\chi(h)\sim1/(h[\ln(1/h)]^{3/2})$. By contrast, weak exchange randomness does not qualitatively alter the susceptibility but has its signature in the heat capacity,  which in the gapless phase is power law in temperature with an exponent dependent on disorder strength.
\end{abstract}


\maketitle

%%%%%%%%%%%%%%%%%%%%%%%%%%%%%%
Disorder in many particle quantum systems is not only unavoidable but also 
an invaluable probe. Isolated impurities can generate new phenomena, as in the Kondo effect \cite{hewson}. They can reveal elusive properties 
of their host, such as the order-parameter symmetry in a superconductor \cite{alloul}. And in some materials a residual defect concentration may dominate low temperature properties \cite{schiffer}. 

Quantum magnets provide a setting for such problems and some of the most challenging questions appear in systems without conventional long range order. Results for the antiferromagnetic spin-half Heisenberg chain illustrate the possibilities that can arise. In the absence of disorder its susceptibility remains finite in the low temperature limit. Site dilution creates free chain ends \cite{Eggert:1992} and leads to a Curie contribution to the susceptibility \cite{sirker-prl}, while disorder in exchange strength generates a random singlet phase \cite{bhatt-lee}, also with a divergent low temperature susceptibility \cite{fisher1994}.  In higher dimensions a single vacancy in a gapped quantum paramagnet again produces a local moment and a Curie-like response \cite{sachdev-science}. Experimentally, local probes such as nuclear magnetic resonance can separate bulk and defect contributions to susceptibility. In particular, studies of some geometrically frustrated magnetic materials that are candidate spin liquids find  substantial defect contributions \cite{schiffer,Olariu:2008}.

Opportunities for an exact treatment of such problems are rare except in one dimension. We show in this paper, however, that the features that make the pure Kitaev honeycomb model \cite{Kitaev:2006ys} tractable  extend to a system with site dilution or exchange randomness.  Without disorder this model provides a solvable instance of a spin liquid, with both gapped and gapless phases, which support fractionalised excitations, abelian and non-abelian \cite{Kitaev:2006ys}. It has extensions to a lattice where the exact ground state is a chiral spin liquid \cite{Yao:2007vn}, to higher spin \cite{baskaran-spin-S}, and to higher dimensions \cite{si-and-yu}. In addition, there are proposals for experimental realisations using cold atoms in an optical lattice \cite{Duan:090402}, and as a solid state system \cite{Jackeli:2009fk}.

The solvability of the Kitaev model rests on the existence of a set of non-dynamical fluxes of an emergent Z$_2$ gauge field. Within each flux sector the Hamiltonian can be reduced to a free fermion problem. Disorder in the form of site dilution or exchange randomness is not an obstacle to this reduction but has dramatic effects on the physical behaviour. 

The simplest form of disorder involves a single {vacancy}. We find that the vacancy binds a flux.  On the sites adjacent to the vacancy a local moment  forms  with a signature in the susceptibility. 
For the pure system the ground-state susceptibility is finite \cite{Kitaev:2006ys}.  In contrast, we show that the vacancy susceptibility is divergent, varying at weak field $h$ in the gapless phase as $\chi(h)\sim \ln(1/h)$.  The moments of different vacancies  interact. Strikingly, two nearby vacancies on the same sublattice have a greatly enhanced susceptibility: $\chi(h) \sim (h[\ln(1/h)]^{3/2})^{-1}$. Our results complement recent work on magnetic impurities  in this model \cite{tripathi}.

We also study the effects of weak randomness in exchange interactions. This type of disorder does not qualitatively influence the susceptibility, which is left finite. It does however change the form of the heat capacity in the gapless phase, since the associated free fermion problem is represented at long wavelengths by a massless Dirac Hamiltonian with a random vector potential. The resulting density of states is power-law in energy \cite{Ludwig:1994fj}, giving a low-temperature heat capacity that is power-law in temperature, with exponents that in both cases are continuously dependent on disorder strength.

\begin{figure}
\epsfig{file=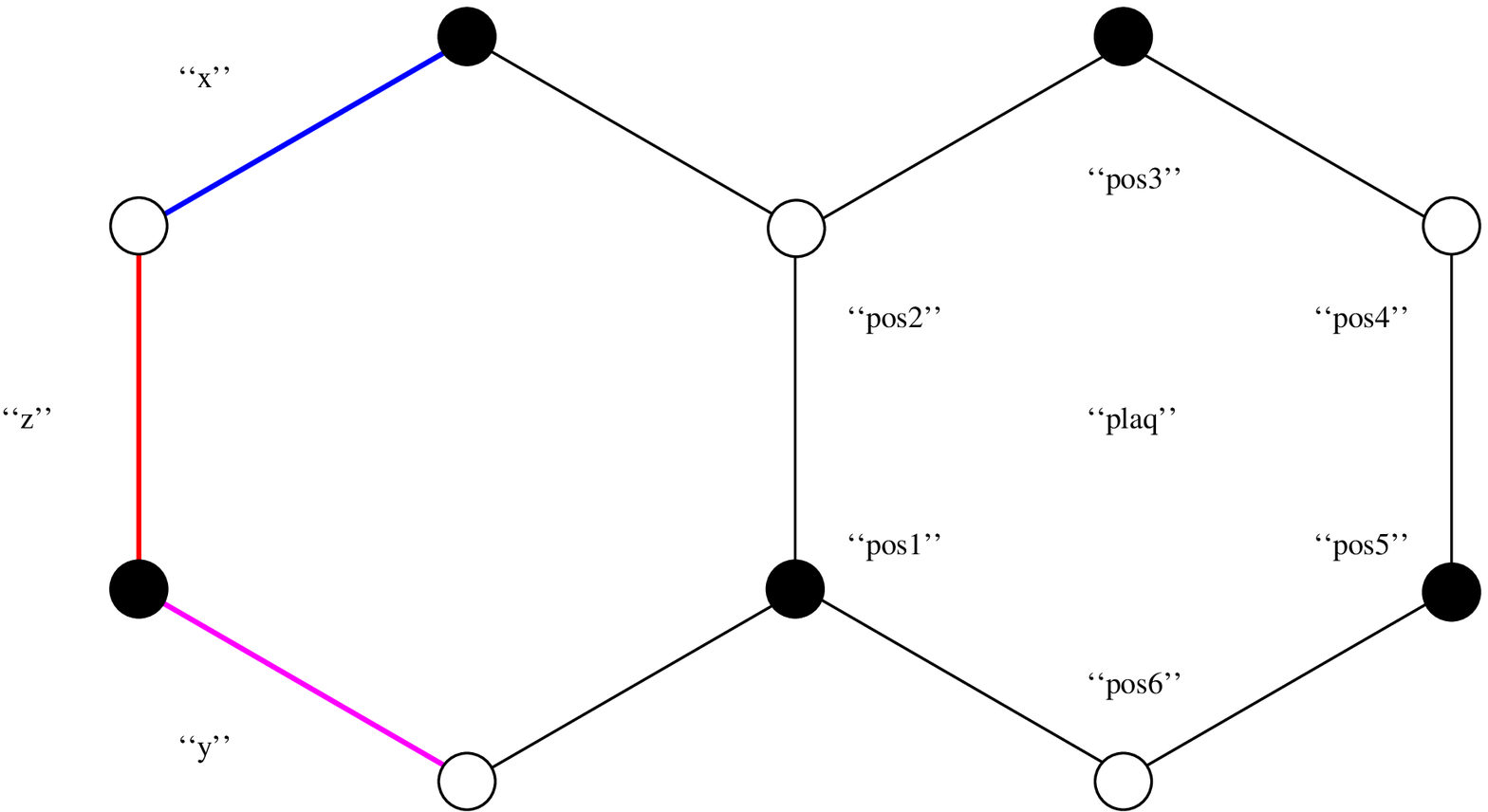, width=3.5cm}\\
\vskip-2mm
\quad(a)\\
\vskip-5mm
\epsfig{file=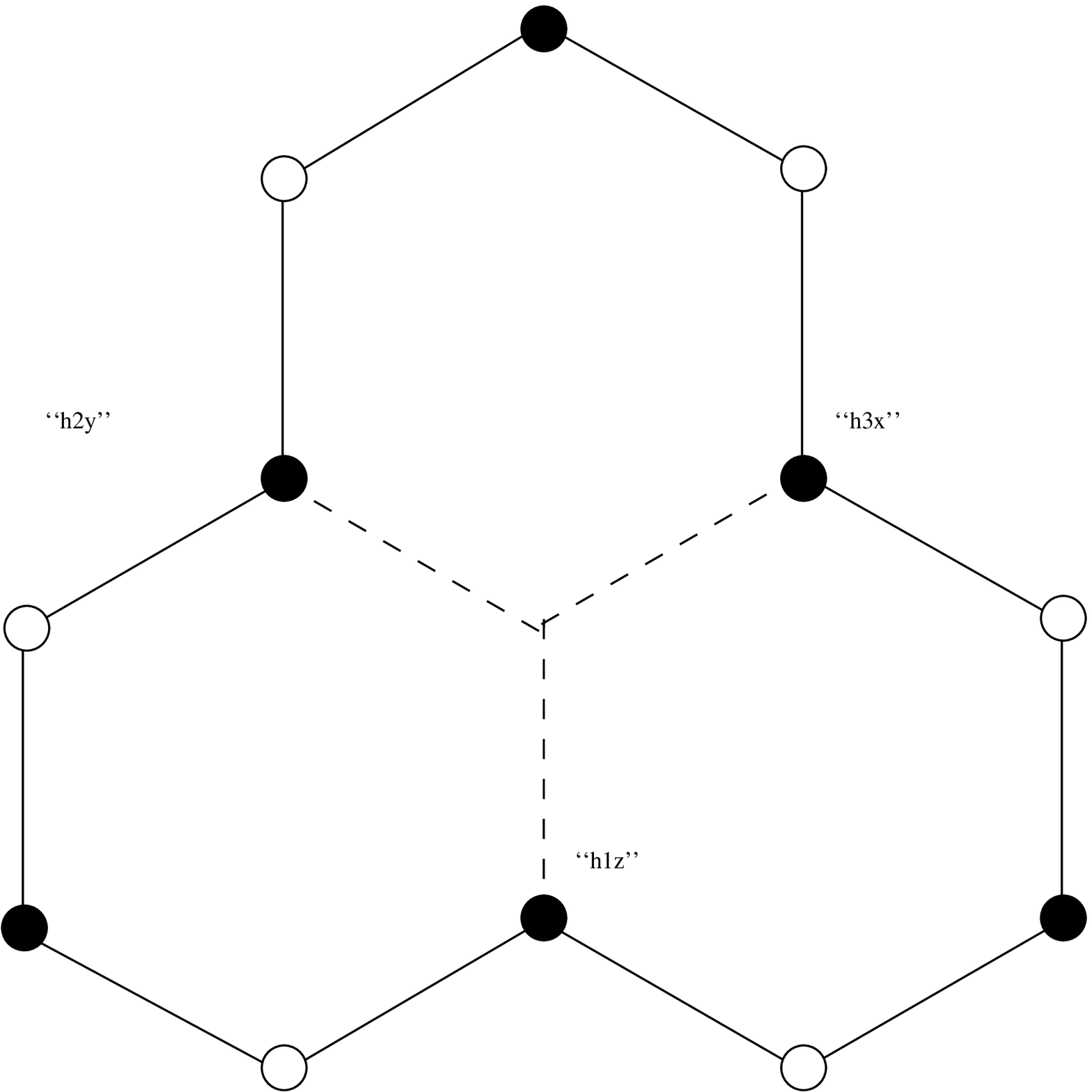, width=3.5cm}\qquad\epsfig{file=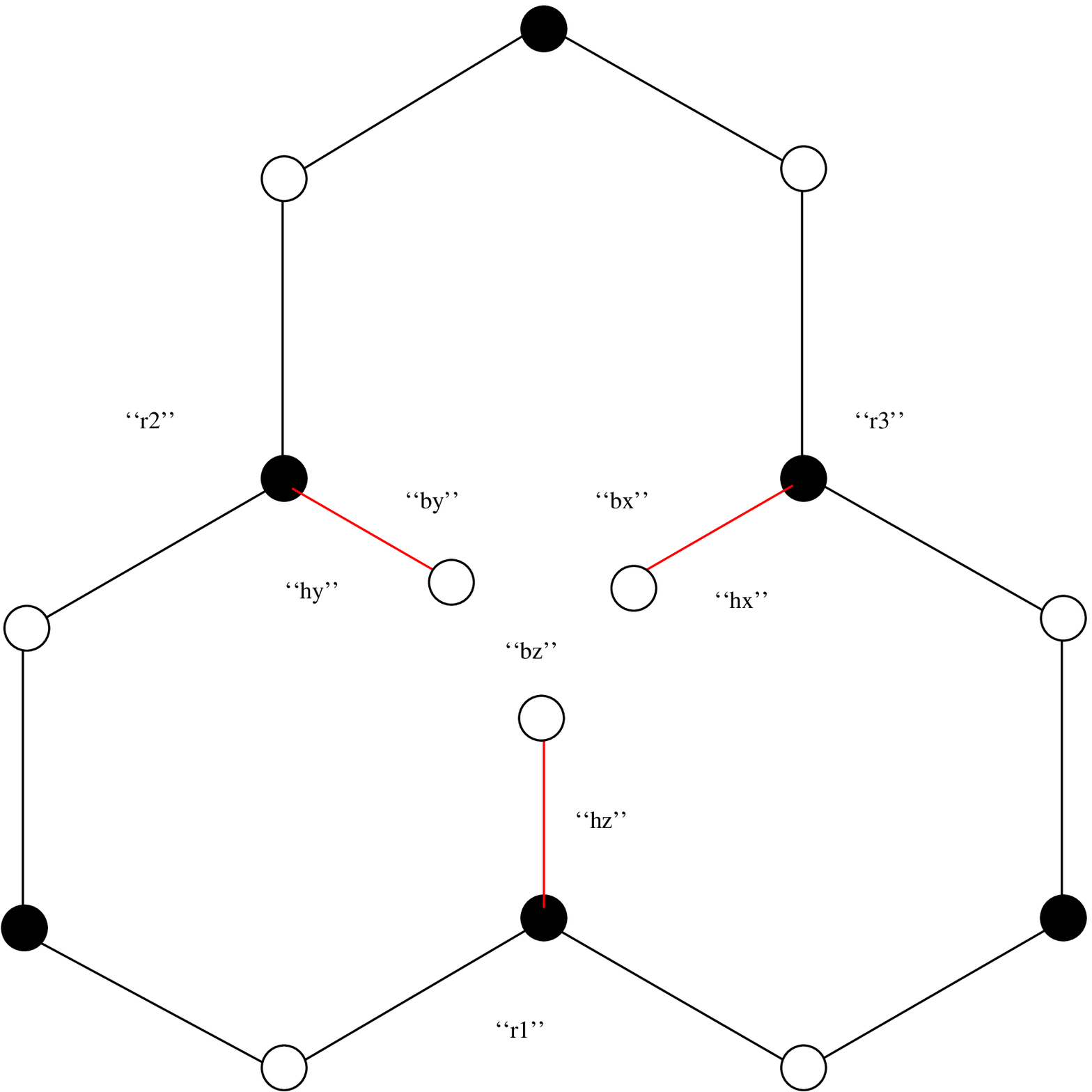, width=3.5cm}\\
(b)\qquad\qquad\qquad\qquad\qquad\qquad(c)
\caption{(a) Labelling $x,y,z$ for bond orientations; site numbering $1\ldots 6$ for plaquette operator $w_{p}$. (b) Sites at which field components $h_x$, $h_y$, and $h_z$ contribute linearly to the projected Hamiltonian with a vacancy.  (c) New sites of the equivalent tight binding model, coupled with hopping $h_x,h_y,h_z$.}
\label{EffHam}
\end{figure}

The Kitaev honeycomb model is defined as follows. A bond on this lattice between nearest neighbour sites $j$ and $k$ takes one of three orientations, labelled using the variable $\alpha_{jk} = x$, $y$ or $z$ as shown in Fig.~\ref{EffHam}(a). A spin one-half variable at site $j$ is represented by the Pauli matrices $\sigma_{j}^{\alpha}$. Exchange of strength $J_{\alpha_{jk}}$ acts between spin components $\alpha_{jk}$ and the Hamiltonian is
\begin{equation}
H = -\dsum_{\langle jk \rangle}J_{\alpha_{jk}}\sigma_{j}^{\alpha_{jk}}\sigma_{k}^{\alpha_{jk}}\,.
\end{equation}
It is exactly soluble, for example, with a local transformation $\sigma^{\alpha} = ib^{\alpha}c$ which represents each spin using the four Majorana fermions $b^x,b^y,b^z$ and $c$ \cite{Kitaev:2006ys}. Then
\begin{equation}
H = \dfrac{i}{2}\dsum_{j,k}J_{\alpha_{jk}}\hat{u}_{jk}c_{j}c_{k}\qquad \hat{u}_{jk} = ib_{j}^{\alpha_{jk}}b_{k}^{\alpha_{jk}}.
\label{quadH}
\end{equation}
The operators $\hat{u}_{jk}$ commute with each other and with $H$.  One can therefore fix the values of $\langle\hat{u}_{jk}\rangle =u_{jk}= \pm1$, move to a subspace of the full Hamiltonian and obtain a bilinear form in the $c_j$'s.  Numbering the sites around a plaquette from 1 to 6 [see Fig.~\ref{EffHam}(a)] the Z$_2$ flux through a plaquette is defined to be $w_{p} = u_{21}u_{23}u_{43}u_{45}u_{65}u_{61}$.  Physical properties of the system depend only on these fluxes \cite{Kitaev:2006ys} but note that many choices of the set $\lbrace u_{jk} \rbrace$ give rise to the same flux sector.  The ground state sector is flux free, e.g. with all $u_{jk} = +1$ for sites $j$ on a particular sublattice \cite{Lieb:1994yq}.  
There is, however, a complication: in transforming to Majorana fermions, the Hilbert space per spin is doubled and projection is necessary to obtain physical states of the system \cite{Kitaev:2006ys}.  Nevertheless, it can be shown \cite{yao-2008} that matrix elements evaluated in a subspace with a fixed set of $\{ u_{jk}\}$ are the same as those obtained using the projected physical states.

In a given flux sector Eq.~(\ref{quadH}) has the form
\begin{equation}
H=\frac{i}{2}\left(\begin{array}{cc}c^{T}_{A}&c^{T}_{B}\end{array}\right)\left(\begin{array}{cc} 0 & M \\ -M^{T} & 0\end{array}\right)\left(\begin{array}{c}c_{A}\\c_{B}\end{array}\right)\,.
\label{hopC}
\end{equation}
Here, for a lattice of $N$ unit cells, $c_{A}$ and $c_{B}$ are each $N$-component vectors  
of Majorana fermion operators, from sublattices $A$ and $B$ respectively, while
$M$ is an $N\times N$ matrix, with entries ${u}_{jk}J_{\alpha_{jk}}$. This Majorana Hamiltonian has the same ground state energy as the complex fermion Hamiltonian 
\begin{equation}
H=i\left(a^{\dagger}_{A}a^{\dagger}_{B}\right)\left(\begin{array}{cc} 0 & M \\ -M^{T} & 0\end{array}\right)\left(\begin{array}{c}a_{A}\\a_{B}\end{array}\right)
\label{hopA}
\end{equation}
at half filling, and we base our study on the associated tight binding model.  In the parameter space with all $J_{\alpha}$ non-negative there are three gapped phases (where one exchange dominates: $J_{z} > J_{x}+J_{y}$ or permutations) and one gapless phase, around the point  $J_{x}{=}J_{y}{=}J_{z}{=}J$ on which we focus below.

To study moment formation due to a vacancy, we introduce a magnetic field $\vec{h}$ with Zeeman coupling to spins. This contribution to the Hamiltonian renders flux dynamical and so spoils solvability \cite{Kitaev:2006ys}. As gaps from the ground state to other flux sectors are ${\cal O}(J)$, at field strength $h\ll J$ one can project onto a given flux sector and work perturbatively in $h$. For the undiluted lattice, matrix elements of the Zeeman energy between states from the same flux sector are all zero and the leading term in a projected perturbation is second order in $h$ \cite{Kitaev:2006ys}.  In the presence of a vacancy, however, the individual fluxes through the surrounding three plaquettes lose physical significance and only their $\rm Z_2$ sum enters the definition of a flux sector. Because of this, the Zeeman energy acquires non-zero matrix elements within a sector, and the first terms in a projected Hamiltonian are now linear in $h$. These terms arise from the specific field components and sites indicated in Fig.~\ref{EffHam}(b): employing the site labelling shown there, the projected Zeeman energy is $H_{\rm Z} = -(h_x \sigma^x_1 + h_y \sigma^y_2 + h_z \sigma^z_3)$.  Written using Majorana fermions, the contributions to $H_{\rm Z}$ have the form $h_{\alpha}\sigma^{\alpha}_{i} = ih_{\alpha} b^{\alpha}_{i}c_{i}$.  They can be represented in the tight binding model,  Eq.~(\ref{hopA}), by the addition of new sites, with the coupling shown in Fig.~\ref{EffHam}(c).  

The task now is to calculate the field-dependence of the ground state energy of the Kitaev model with a vacancy, using the tight binding model of Fig.~\ref{EffHam}(c).
Consider first the behaviour in a gapped phase.
A consequence of the bipartite structure of this tight binding model, in any flux sector, is that 
finite energy eigenstates appear only as positive and negative energy pairs. There may in addition be zero energy states.  At ${\bf h}=0$ there are four of these, but if all components of $\bf h$ and all $J_\alpha$ are non-zero there are only two such states, accompanied by a finite energy pair with a separation that is linear in $h$ for small $h$. As a result, the ground state energy decreases linearly with $h$, reflecting a finite vacancy moment which varies continuously with the exchange parameters  at vanishing $h$ in a gapped phase.

Vacancy properties in the gapless phase are more subtle, and cannot be found by considering only a finite number of energy levels. 
We instead proceed by computing the Green function $G({\bf h})$ for this tight binding model. It is a function of a complex energy parameter $z$. Defining $\mathcal{E}(z,h)=z\text{Tr}\left[G({\bf h})-G({\bf 0})\right]$, we evaluate the change in ground state energy within the flux sector for the Kitaev model with a vacancy and field, compared to its zero field value, as  an anti-clockwise integral around the negative real axis:
\begin{equation}\label{energy-integral}
\mathcal{E}(h) = \dfrac{1}{2\pi i}\oint\mathcal{E}(z,h)dz\ .
\end{equation}
 As advertised above, the presence of a vacancy  changes the flux sector in which the ground state lies. 
 However, we postpone discussion of this aspect and focus initially on the simpler behaviour in the flux free sector, which we will make use of for the vacancy pair calculation below. 

In essence, our calculation of  $\mathcal{E}(z,h)$ depends on the fact that  $G({\bf h})$ is 
related by a finite rank perturbation to the Green function $G_0$ for the undiluted lattice, which is known \cite{horiguchi:1411}. Moreover, at small $h$ the integral in Eq.~(\ref{energy-integral}) is dominated by contributions from small $z$, and so results are governed by the behaviour of $G_0$ for $|z|\ll J$. The standard T-matrix formalism casts $G({\bf h})$ in terms of the matrix elements $G_0({\bf r},z)$ of $G_0$ between sites with separation $\bf r$. These enter in the combination $g(z) = G_{0}(0,z)+G_{0}(1,z)^{2}/G_{0}(0,z)$, which is the site-diagonal element of $G({\bf 0})$ at one of the sites (e.g. $r_1$) adjacent to the vacancy. Expressions are simplest if only one component of $\bf h$ is non-zero, and in this case
\begin{equation}
\mathcal{E}(z,h) = h^2[g(z)-z\partial_zg(z)]/[z-h^2g(z)]\,.
\end{equation}
Setting $J{=}1$, one has for small $z$ \cite{horiguchi:1411} 
\begin{equation}\label{GreensFn}
G_{0}(0,z) \sim \lambda z\ln\left[-(\mu z)^2\right] \qquad
G_{0}(1,z) \sim i\nu
\end{equation}

with $\lambda = 1/\sqrt{3} \pi$ and $\mu = \nu = 1/3$ .  The $z$-dependences of Eq.(\ref{GreensFn}) are a direct consequence of the massless Dirac spectrum for the nearest neighbour tight binding model on the honeycomb lattice, and hold with appropriate values for $\lambda$, $\mu$ and $\nu$ throughout the gapless phase. 

In this way we obtain the asymptotic behaviour for small $h$ of the vacancy energy $\mathcal{E}(h) \sim -h \nu[2\lambda\ln(1/h)]^{-1/2}$ and magnetisation 
\begin{equation}
m(h)= -\partial_{h}\mathcal{E}(\hh)\sim \nu [2\lambda \ln(1/h)]^{-1/2}\,.
\label{EandM}
\end{equation}
It is apparent from the form of the projected Zeeman energy $H_{\rm Z}$ that this magnetisation is entirely localized on sites adjacent to the vacancy. Strikingly, each of these sites, labelled $r_{1}, r_{2}, r_{3}$ in Fig.~\ref{EffHam}(c), carries a separate component ($m_{x}, m_{y}$ and $m_{z}$, respectively) with relative magnitude proportional to the corresponding component of $\bf h$.  From Eq.~(\ref{EandM}) we find the defect susceptibility $\chi(h)=\partial_h m(h)$ given above, which diverges  in the small  $h$ limit. Thermal fluctuations at a temperature $T$ small compared to the exchange can be treated as fermion excitations within one flux sector, giving a linear  vacancy susceptibility that diverges as $1/[T\ln(1/T)]$.

\begin{figure}[ht]
\epsfig{file = 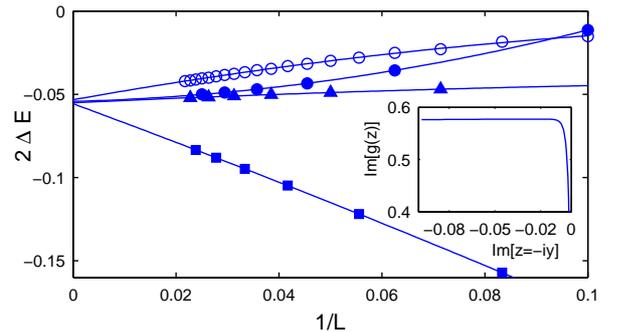, width=8cm, height=4.5cm}
\caption{For systems with two vacancies, the ground state energy difference $2\Delta E$ between the sector with fluxes on vacancies and the flux free sector, as a function of inverse linear system size $1/L$.  Open symbols: systems with open boundary conditions. Filled symbols: systems with periodic boundary conditions, for which there are three classes of behaviour, determined by $L$ mod 3.  Lines are quadratic fits in $1/L$. Inset:  $g(z)$ in the presence of a flux pair separated by  $d=170$.}
\label{FluxEnergy}
\end{figure}

%%%%%%%%%%%%%%%%%%%%%%%%%%%%%%%%%%%%%%%%%%%%
So far we have discussed response in the flux free sector.  In fact, we find that a vacancy has a flux bound to it in the ground state. Deep in a gapped phase this can be demonstrated analytically. Consider $J_x, J_y \ll J_z$. For $J_x{=}J_y{=}0$ the ground state consists of paired spins on $z$ bonds and has degeneracy $2^N$. Lifting of this degeneracy can be studied using $J_x/J_z$ and $J_y/J_z$ as small parameters. One obtains an effective Hamiltonian with leading  terms proportional to the flux through each elementary plaquette of the lattice \cite{Kitaev:2006ys}. It is minimised by taking zero flux through all hexagonal plaquettes but $\pi$ flux through the vacancy plaquette. In the gapless phase this leading order calculation is insufficient. Instead we compute the energy of different flux sectors numerically: a flux binding energy of $-0.027J$ is demonstrated by the negative intercept in Fig.~\ref{FluxEnergy}. It would be intriguing to find whether mobile vacancies also bind flux, which would have implications for their relative statistics.

With this conclusion in mind, we now revisit the calculation of vacancy magnetisation. The result we have presented for the flux free sector hinges on the behaviour of $g(z)$. To carry out a similar calculation for a vacancy with a flux attached, we require the elements of the Green function for an undiluted honeycomb lattice with $\pi$ flux through one hexagonal plaquette, evaluated between sites lying on this hexagon. We find numerically that the flux generates quasi-localised modes at low energies, which in turn remove the singular behaviour of $g(z)$ at small $z$. This results in a ground state vacancy magnetisation $m(h) \sim h\ln(1/h)$ and a low temperature linear vacancy susceptibility that diverges as $\ln(1/T)$.

In detail, an efficient computation of Green function elements is possible for an infinite lattice with two fluxes through hexagons separated by a distance $d$, for $d\lesssim 500$. Results of such a calculation are displayed in the inset to Fig.~\ref{FluxEnergy}, where $\Im[g(z)]$ is shown as a function of $z$, taken on the imaginary axis.  It is a constant for $1\gg |z| \gg 1/d$, but sensitive to the finite flux separation for $1/d \gg |z|$.  The behaviour of the vacancy magnetisation with a bound flux is a consequence of constant $g(z)$ at small $z$.

We next consider a pair of vacancies, represented by a tight binding model with two defect centres of the type shown in Fig.~\ref{EffHam}(c). We find that there is an interaction between the magnetic moments formed around each vacancy, which is weak if the pair separation is large. The interaction sets a field scale. Above it, the susceptibility is a sum of two independent vacancy contributions. Below it, behaviour depends on the relative sublattices occupied by the two defects. Moreover, since the weak-field response involves low-energy and long-distance features of the system, it is controlled by the ${\rm Z}_2$ sum of the fluxes associated with the vacancies, which is zero. Results for the flux-free sector, expressible in terms of $G_0$, therefore also illustrate behaviour in the ground state sector where vacancies bind fluxes.

The discussion centred on Eq.~(\ref{energy-integral}) can be repeated for the two-vacancy problem, with the corresponding Green function expressed in terms of $G_0$, but one further matrix element is required in addition to the two in Eq.~(\ref{GreensFn}): the one between the two vacancy sites. This enters the generalisation of 
$\mathcal{E}(h)$ and sets a scale for $h$. With vacancies on opposite sublattices and $z\ll |\dd|^{-1}$ its value is
\begin{equation}
G_{0}(\dd,z) \sim |\dd|^{-1}{\sin(\KK.{\dd}-\theta)}\,,
\end{equation}
where $\KK = (2\pi/3, -2\pi/3)$ is the momentum at one of the Dirac points and $\theta$ is the angle between $\dd$ and the $x$ axis \cite{wang:125417}.  For $h\ll h_{\text c}(\dd) =| G_{0}(\dd,0) | (\ln[|1/G_{0}(\dd,0)|])^{-1/2}$ the vacancy susceptibility is large but field independent, while in the opposite limit ($h_{\text c}(\dd)  \ll h \ll 1$) vacancies are independent. With both vacancies on the same sublattice, $G_{0}(\dd,z)$ vanishes as $z$ approaches zero, and Zeeman energy dominates over the interaction between vacancies at all $h$. This has the remarkable consequence that a pair of vacancies on the same sublattice has a {\em parametrically} larger weak-field susceptibility ($\chi(h) \sim (h[\ln(1/h]^{3/2})^{-1}$) 
than an isolated vacancy with bound flux.

We now turn to a discussion of the influence of disorder in the strength of exchange interactions. Small amplitude disorder in the absence of vacancies has no qualitative effect on some aspects of behaviour: the ground state has finite susceptibility and is in the flux free sector. Exchange randomness does however enter the free fermion description of states within this sector. We recall that without disorder this description reduces at low energy to two copies of the Dirac Hamiltonian, which in the gapless phase is massless \cite{Kitaev:2006ys}. Randomness in exchange interactions appears in the Dirac Hamiltonian as a random vector potential. The consequences of such disorder for fermion eigenstates have been studied extensively   \cite{Ludwig:1994fj}. Most importantly in our context, the fermion density of states is proportional to a power of energy, with an exponent that depends on disorder strength.  We introduce weak, smooth exchange disorder of the form $J_{\alpha}(\rr) = J(1+\epsilon_{\alpha}(\rr))$, with correlations $\langle \epsilon_{\alpha}({\bf r}) \epsilon_{\beta}({\bf r}')\rangle =\frac{8}{9}\pi\Delta  \delta_{\alpha\beta}f({\bf r} - {\bf r}')$, where $\sum_{\bf r} f({\bf r})=1$ and $f({\bf r})$ decreases with smoothly with ${\bf r}$. Then for small
$\Delta$
 the heat capacity $C$ has the low temperature form $C\sim T^{2/(1+\Delta)}$. Note that this implies that -- especially in a system with strong spin-lattice coupling -- observing $C\propto T^2$ is not an experimental requirement for identifying a magnet with a gapless `Dirac' excitation spectrum.  

In summary, we have studied the Kitaev honeycomb model as an example of a spin liquid that is solvable even in the presence of disorder. At an isolated vacancy a flux is bound and a local moment forms, with singular susceptibility in the gapless phase. Weak exchange disorder does not influence the susceptibility but leads to a continuously variable heat capacity exponent. 
Finally, as indicated by the rich and intriguing behaviour  of vacancy pairs, we note that the properties of the model with a finite density  of vacancies represents an intriguing and demanding open problem. 

This work was supported in part by EPSRC under Grant No. EP/D050952/1.

\end{document}